\documentclass[amsmath,amssymb,showpacs,reprint,
floatfix,aps,prl,superscriptaddress]{revtex4-1}
\usepackage[hidelinks,colorlinks=true,allcolors=blue]{hyperref}
\usepackage{amssymb}
\usepackage{amsmath}
\usepackage{graphicx}
\usepackage{physics}
\usepackage{braket}
\usepackage{xcolor}
\usepackage[labelfont=bf]{caption}
\usepackage{subcaption}
\usepackage{multirow}
\graphicspath{ {./}{figures/}{figures/tikz/}}

\renewcommand{\vec}[1]{\mathbf{#1}}

\begin{document}
\setlength{\intextsep}{5pt}
\title{ Renormalized quasiparticles, topological monopoles and superconducting line nodes in heavy fermion CeTX$_3$ compounds}
\author{Vsevolod Ivanov}
\email[E-mail: ]{vivanov@ucdavis.edu}
\affiliation{%
Department of Physics, University of California, Davis, California 95616, USA}
\author{Xiangang Wan}
\affiliation{%
Department of Physics, Nanjing University, Nanjing 210093, China}
\author{Sergey Y. Savrasov}
\email[E-mail: ]{savrasov@physics.ucdavis.edu}
\affiliation{%
Department of Physics, University of California, Davis, California 95616, USA}
\vspace{-5pt}
\begin{abstract}
\noindent
Non-centrosymmetric superconductors have recently attracted much attention, 
since the lack of inversion symmetry mixes spin singlet and triplet pairing 
states, which may allow the realization of topological superconductivity. In 
this work, we study the electronic properties of the family of inversion-broken
CeTX$_3$ heavy-fermion superconductors, finding topological nodal lines as well
as Dirac and Weyl points, which are renormalized closer to the Fermi energy by
correlations. We find that the Weyl nodal lines have a substantial effect on 
the Fermi surface spin structure of the normal state, and lead to line nodes in the superconducting phase.
%
\end{abstract}
\maketitle
%
%
%
%
\indent
Superconductivity (SC) in non-centrosymmetric compounds has received much 
attention due to their potential for hosting unconventional pairing states. The
lack of inversion symmetry permits antisymmetric spin-orbit coupling (ASOC) which
splits the Fermi surface (FS) and mixes spin-singlet and spin-triplet SC
pairing states.\\
%
%
%
%
\indent The CeTX$_3$ (T = Co, Rh, Ir, X = Si, Ge) family of compounds 
crystallize in the BaNiSn$_3$-type structure ($I4mm$ space group no. 107), 
which breaks spatial inversion symmetry. With the exception of paramagnetic 
CeCoSi$_3$, their low-temperature phases are anti-ferromagnetic (AF) at ambient 
pressure. Application of pressure suppresses the N\'eel temperature to zero, 
where the magnetic ground state gives way to SC. The SC in this group exhibits 
many unconventional features, including upper critical fields $H_{cw}$ that far 
exceed the Pauli limiting field $H_p[T] \sim 1.86T_C[K]$\cite{cecoge3-SC, 
cerhsi3-SC,ceirge3-SC,ceirsi3-SC,cerhge3-SC,cetx3-SC,ceirsi3-SC2,cerhsi3-SC2} 
which has been suggested as evidence of an odd parity SC gap function. Recent 
works argue that AF fluctuations play a role in the development of SC,
indicating the importance of the spin structure to the unconventional physics 
in these compounds.\\
\indent
The absence of inversion symmetry is also a necessary ingredient for the
existence of topological Weyl points. Since the role of ASOC and lack of
inversion symmetry in the development of the SC state is not well understood,
we hope to shed some light by investigating the topological properties of these
materials. Furthermore, the narrow Ce-$4f$ band is sensitive to 
temperature and pressure, allowing Weyl points to be tuned without introduction 
of chemical or site disorder. This feature makes these heavy fermion materials 
promising candidates for the study of Weyl physics \cite{QSi} in the proximity 
of SC and quantum criticality.\\
%
%
%
%
\indent
Our electronic structure calculations are performed within the framework of 
the full potential linear muffin–tin orbital method with spin-orbit coupling, 
using the experimentally measured lattice parameters \cite{2008cetx3, 
cerhge3-struct, cecoge3-struct, cerhsi3-fs}. The compounds are locked to the 
paramagnetic state to mimic the experimentally observed suppression of 
magnetism by pressure. The on-site interactions between the Ce-$4f$ electrons 
must be treated with special care, as the strong Coulomb repulsion narrows the 
bandwidth considerably. We handle renormalization of quasiparticle bands 
through the LDA+Gutzwiller (LDA+G) method, taking Hubbard $U$ values of $5$eV 
and $6$eV\cite{Ce-U}. The method is described in more detail in 
Refs. \cite{sav-ce-guz,guz-dft,lda+guz,lda+guz2}\\
\indent
In LDA+G, the double-counting potential must carefully be chosen to account for 
the Coulomb correction included in both the single-particle and interacting 
terms of the Hamiltonian. Specifically, for the electron self-energy correction,
$\Sigma_\alpha(0) - V_{\text{DC},\alpha}$, there are several options for the
double counting potential $V_{\text{DC},\alpha}$ \cite{sav-ce-guz}. One such
option is to set $V_{\text{DC},\alpha} = \Sigma_\alpha(0)$, which leaves the
LDA FS intact. Another option is to compute the crystal-field modifications
self-consistently using an average over orbital self-energies, $V_{DC,\alpha}
= \frac{1}{N} \sum_{\alpha}^N \Sigma_\alpha(0)$.
\indent
For the CeTX$_3$ compounds, the crystalline electric field (CEF) effect of the
tetragonal symmetry lifts the degeneracy of the J=5/2 total angular momentum
state, splitting it into three doublets. Magnetic susceptibility and inelastic
neutron scattering experiments \cite{cerhge3-cef, ceirsi3-cef, cecoge3-cef,
ceirge3-cef} have determined the ground state doublet to be $\Gamma_7^{(1)}$
with $\Gamma_6$ and $\Gamma_7^{(2)}$ slightly higher in energy. Our LDA
calculation shows that the lowest energy doublet hybridizes with the four bands
crossing the Fermi energy ($E_F$), which are largely responsible for the shape
of the FS. This is consistent with prior works that show qualitative agreement
between the LDA FS and experimental de Haas-van Alphen measurements for
CeRhSi$_3$ \cite{cerhsi3-fs}. In order to best match the experimentally
determined Fermi surfaces and mass enhancements we take a phenomenological
approach, selecting a hybrid double counting scheme which independently treats 
the lowest energy doublets while the remaining states are shifted upward by 
0.1Ry \cite{sup}. A different choice of shift parameter does not affect the
states near the Fermi energy, and does not change the conclusions of our work. 
An analogous energy shift was used to find the FS of the 
isostructural LaTX$_3$, which is presumed to be very similar to that of the 
respective CeTX$_3$ compounds since their Ce-$4f$ electrons are highly 
localized \cite{cecoge3-fs, cerhsi3-cef}. \\
%
%
%
%
%
%
%
%
%
\begin{table}[b]
	\vspace{-16pt}
	\caption{Quasiparticle residues $z_\alpha$ for the lowest energy 
		states for the members of the CeTX$_3$ series.}
	\begin{tabular}{l c c c c c c }
		\hline\hline
		& \multicolumn{3}{c}{$z_{\text{LDA+G}}$ ($U$=5 eV)} 
		& \multicolumn{3}{c}{$z_{\text{LDA+G}}$ ($U$=6 eV)} \\
		& $\Gamma_7^{(1)}$ & $\Gamma_6$ & $\Gamma_7^{(2)}$ 
		& $\Gamma_7^{(1)}$ & $\Gamma_6$ & $\Gamma_7^{(2)}$ \\
		\hline
		CeCoSi$_3$ 	& 0.59	& 0.57	& 0.87	& 0.54	& 0.52	& 0.82	\\
		CeRhSi$_3$ 	& 0.43	& 0.41	& 0.86	& 0.37	& 0.36	& 0.81	\\
		CeIrSi$_3$ 	& 0.43	& 0.42	& 0.86	& 0.38	& 0.36	& 0.81	\\
		CeCoGe$_3$ 	& 0.38	& 0.36	& 0.85	& 0.33	& 0.32	& 0.78	\\
		CeRhGe$_3$ 	& 0.16	& 0.14	& 0.92	& 0.12	& 0.10	& 0.89	\\
		CeIrGe$_3$ 	& 0.15	& 0.14	& 0.93	& 0.11	& 0.09	& 0.91	\\
		\hline\hline
	\end{tabular}
	\label{gammas}
\end{table} 
%
%
\indent
Our LDA+G procedure yields band-dependent quasiparticle residues $z_\alpha$, 
which are summarized in Table \ref{gammas}. It is worth noting that the 
$\Gamma_7^{(2)}$ doublet has been determined to be the lowest lying state in 
CeRhSi$_3$ \cite{cerhsi3-cef}. However, our calculations place the 
$\Gamma_7^{(1)}$ doublet at the lowest energy for all six 
isoelectronic compounds.
\begin{figure}[b]
	\vspace{-16pt}
	\includegraphics[width=.50\textwidth]{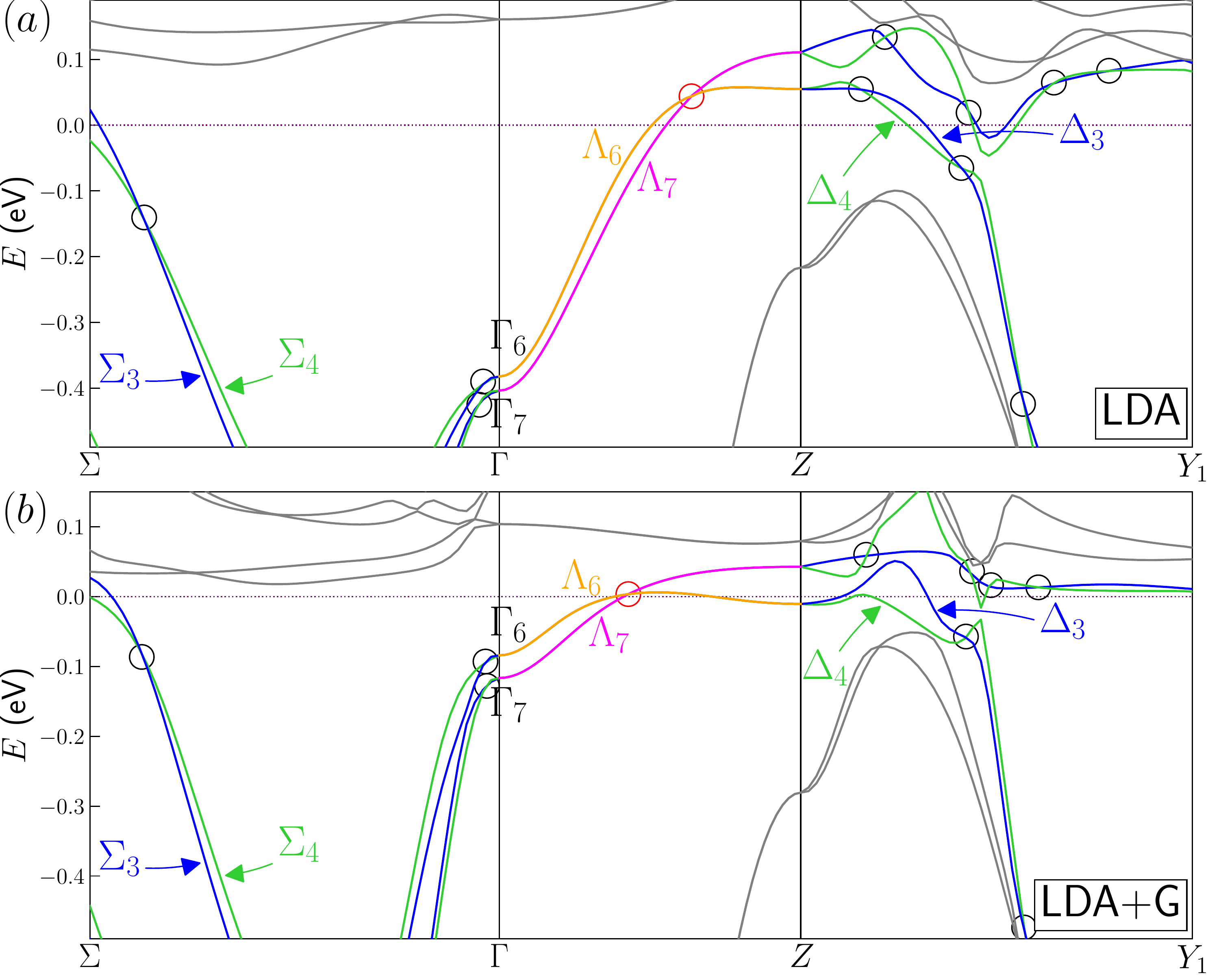}
	\caption{Band structures for CeCoGe$_3$ using (a) LDA, and (b) LDA+G. Bands
		are labeled with their character representations according to their
		mirror eigenvalue: $\Sigma_3$/$\Delta_3$ (blue) for -$i$ and 
		$\Sigma_4$/$\Delta_4$ (green) for $i$ within the mirror planes. Along 
		the $\Gamma-Z$ line, doublets $\Lambda_6$ (orange) and $\Lambda_7$ 
		(magenta) form a DP. NL crossings and DPs are indicated by black and 
		red circles respectively.}
	\label{chars}
\end{figure}
\indent
The trends in the CeTX$_3$ series can be understood in terms of a Doniach phase
diagram arising from competing RKKY and Kondo interactions \cite{Doniach1977}.
The tuning parameter in the Doniach phase diagram is $\abs{J_{cf}}N(0)$ where
$J_{cf}$ is the magnetic exchange interaction and $N(0)$ is the density of
states at the $E_F$. Experimentally, this parameter can be tuned by
compressing the lattice using pressure, resulting in a greater hybridization of
the conduction and Ce-$4f$ bands, thus decreasing the localization of the
electrons. This is reflected directly in the trend of N\'eel
temperatures, with CeTGe$_3$ compounds exceeding their Si counterparts, 
($T_N$ = 21K, 14.6K, 8.7K vs. 0K, 1.8K, 5.0K for T = Co, Rh, Ir),
due to their larger lattice constants \cite{2008cetx3}.
The N\'eel temperatures of CeRhSi$_3$ and CeIrSi$_3$ are suppressed to zero at
relatively low pressures $P_c \sim 2$ GPa, indicating their proximity to a
quantum critical point. \\
%
\begin{figure}[ht]
	\includegraphics[width=0.48\textwidth]{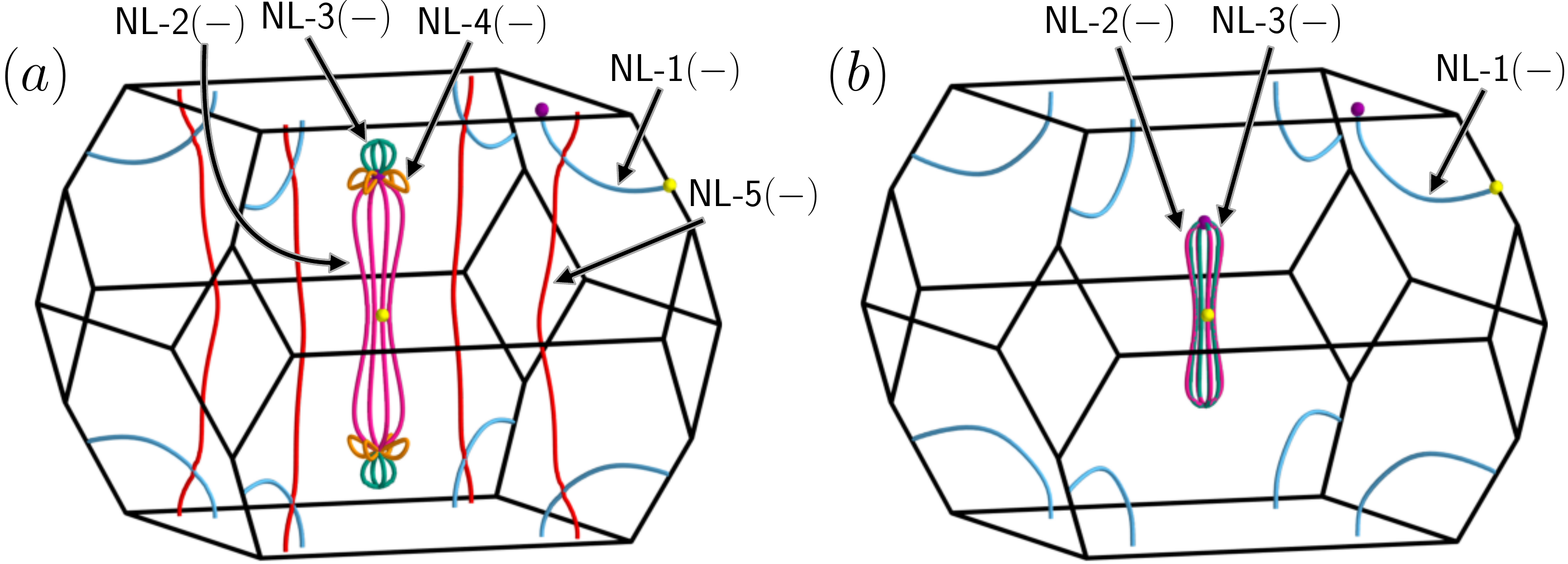}
	\caption{Selected NLs in CeCoGe$_3$ for LDA (a), and LDA+G (b), with colored
		spheres showing start (yellow)/end (purple) points for NLs plotted in Fig. \ref{renorm}.}
	\label{nl}
\end{figure}
\indent 
The computed $z_\alpha$ values follow a decreasing trend with increasing 
lattice volumes, and qualitatively match the experimental trend of larger 
quasiparticle masses as the mass of the transition metal atom increases. 
These imply a factor of $\sim$2$-$9-fold increase in Sommerfield $\gamma$ 
values, but experimental measurements on CeTX$_3$ compounds in the high 
pressure paramagnetic state are not presently available for comparison.\\
%
%
\indent
The bands crossing $E_F$ are predominantly Ce-$4f$ in character, 
with a minor contribution from the transition metal $d$-orbitals away from 
the Fermi level. When Coulomb interactions are considered through the LDA+G 
calculation described above, their bandwidth is narrowed and the Fermi level 
is pinned to the lower doublet due to the increased density of states (Figure 
\ref{chars}), changing the electronic structure and associated topological 
features near $E_F$. We emphasize that while the particular number 
and shape of the topological features depend on the choice of double counting 
potential and magnitude of Hubbard-$U$, their existence is guaranteed by 
symmetry and robust to correlations. Since the CeTX$_3$ compounds are 
isoelectronic, the general picture of their topological properties is the same,
with each compound hosting different sets of particular features based on the 
relative band positions determined by the CEF splitting. For the remainder of 
this work we will focus on describing the electronic properties of CeCoGe$_3$,
which hosts representative members of each type of topological features found
in the series, including 
Dirac points (DPs) \cite{dirac1,dirac2,cd3as2,dirac4,dirac5,dirac6}, 
Weyl points (WPs) \cite{weyl1-sergey,weyl2,weyl3,weyl4}, 
and nodal lines (NLs) \cite{nodal1,nodal2,nodal3,nodal4}. \\
%
%
%
\begin{table}[b]
	\vspace{-16pt}
	\caption{Non-equivalent WPs of CeCoGe$_3$, with columns: topological charge 
		(C), number of symmetry equivalent WP in this set (\#), location
		($\vec{k}_{\text{Weyl}}$) given in units of $(2\pi/a,2\pi/a,2\pi/c)$,
		and energy in meV (E). The Fermi energy is set to 0 eV.}
	\hspace*{-0.4cm}
	\begin{tabular}{c  c | c  c | c  c }
		\hline\hline
		\multicolumn{2}{c|}{CeCoGe$_3$} & \multicolumn{2}{c|}{ LDA} 
		& \multicolumn{2}{c}{LDA+G}\\
		 C & \#	&	$\vec{k}_{\text{Weyl}}$	& E & $\vec{k}_{\text{Weyl}}$ & E\\
		\hline
		$+1$&8	& (0.097, 0.187, 1.000)	& -109 & (0.161, 0.133 , 1.000)	&  -49\\
		$-1$&16	& (0.118, 0.152, 0.556)	& -140 & (0.131, 0.168 , 0.586)	& -110\\
		$+1$&16	& (0.235, 0.271, 0.676)	&  +78 & (0.167, 0.236 , 0.530)	&  +33\\
		$-1$&16	& (0.057, 0.285, 0.996)	& +118 & (0.083, 0.221 , 0.617)	&  +37\\
		\hline\hline
	\end{tabular}
	\label{weyl-table}
\end{table}
%
%
\indent
To locate and confirm the topological features, we use a one shot method for 
data mining the bands \cite{mmm}. We divide the BZ into an initial $20 \times 
20 \times 20$ $\vec{k}$-grid, computing the integral of Berry curvature fluxes 
through the surface of each k-cube to find sources and sinks. The locations of 
these topological points are recursively refined by repeating the procedure on 
a $4 \times 4 \times 4$ grid within their $\vec{k}$-cube until the desired 
precision is achieved, thus resolving much finer details of the material topology.\\
\indent
We find two classes of WPs in CeCoGe$_3$. The first appears in sets of eight 
confined to the $k_z=0$ plane, while the second comes in sets of 16 which are 
additionally separated in the $k_z$ direction. Table \ref{weyl-table} shows
selected WPs of CeCoGe$_3$ listed along with their presumptive
counterparts in LDA+G, which are shifted slightly in momentum space due to band
renormalization. In total, CeCoGe$_3$ has seven (eight) non-equivalent Weyl
points in LDA (LDA+G); additional details can be found in the Supplemental Material (SM) \cite{sup}.\\
%
%
%
%
%
\begin{figure}[b]
	\vspace{-20pt}
	\includegraphics[width=.48\textwidth]{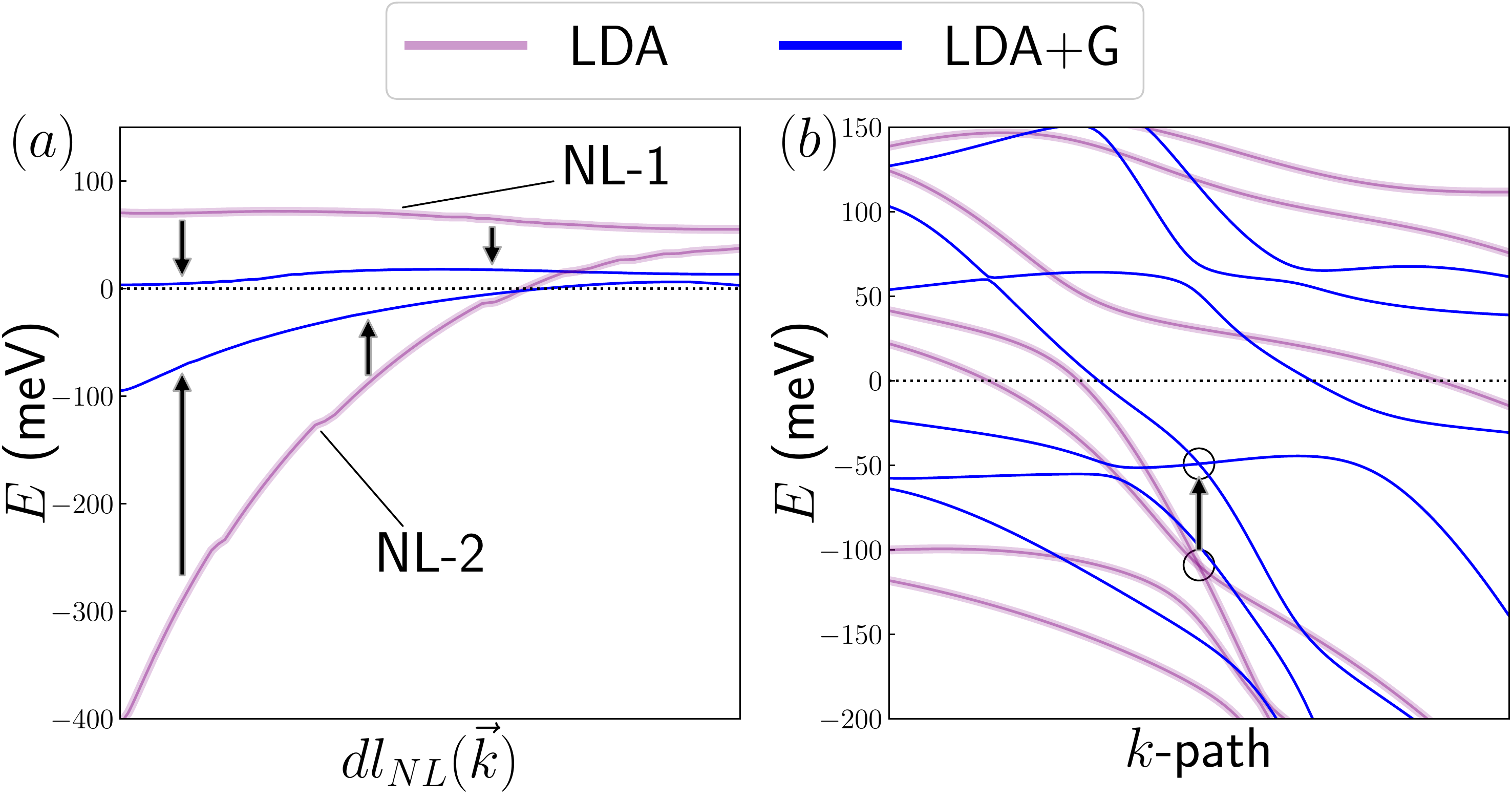}
	\caption{Renormalization of topological features in CeCoGe$_3$ between LDA
		(violet) and LDA+G (blue). (a) Renormalization of NL-1 and NL-2. Energy
		is plotted along the length of each NL (normalized to unity), with
		start/end points as shown in Fig \ref{nl}. (b) Plots of bands around
		the first WP from Table \ref{weyl-table}. $k$-path is the straight line
		connecting $\vec{k}_{\text{Weyl}} \pm 0.1 \hat{k}_y$.}
	\label{renorm}
\end{figure}
\indent
The most striking topological structure in the BZ is the set of nodal lines 
emerging from the Dirac point in this material. The band inversion mechanism 
generating the DP along the $\Gamma-Z$ axis is similar to that 
responsible for the DP in the inversion broken Cd$_3$As$_2$ 
\cite{cd3as2}, which shares the C$_{4v}$ point group symmetry. Along the 
$\Gamma-Z$ direction, compatibility relations for the double group connect 
$\Gamma_7 \rightarrow \Lambda_7$ and $\Gamma_6 \rightarrow \Lambda_6$. When 
moving along $\Gamma-Z$, the lowest lying 
$\Lambda_7$ Kramer's doublet switches with the $\Lambda_6$ doublet. The DP
formed by the two doublets persists with the inclusion of band
renormalizations, shifting from a position $k_z = 0.644\frac{2\pi}{c}$ in LDA
to $k_z = 0.4285\frac{2\pi}{c}$ in LDA+G, closer to the $\Gamma$ point, as
shown in Figure \ref{chars}.\\
%
\begin{figure*}[htb]
	\includegraphics[width=\textwidth]{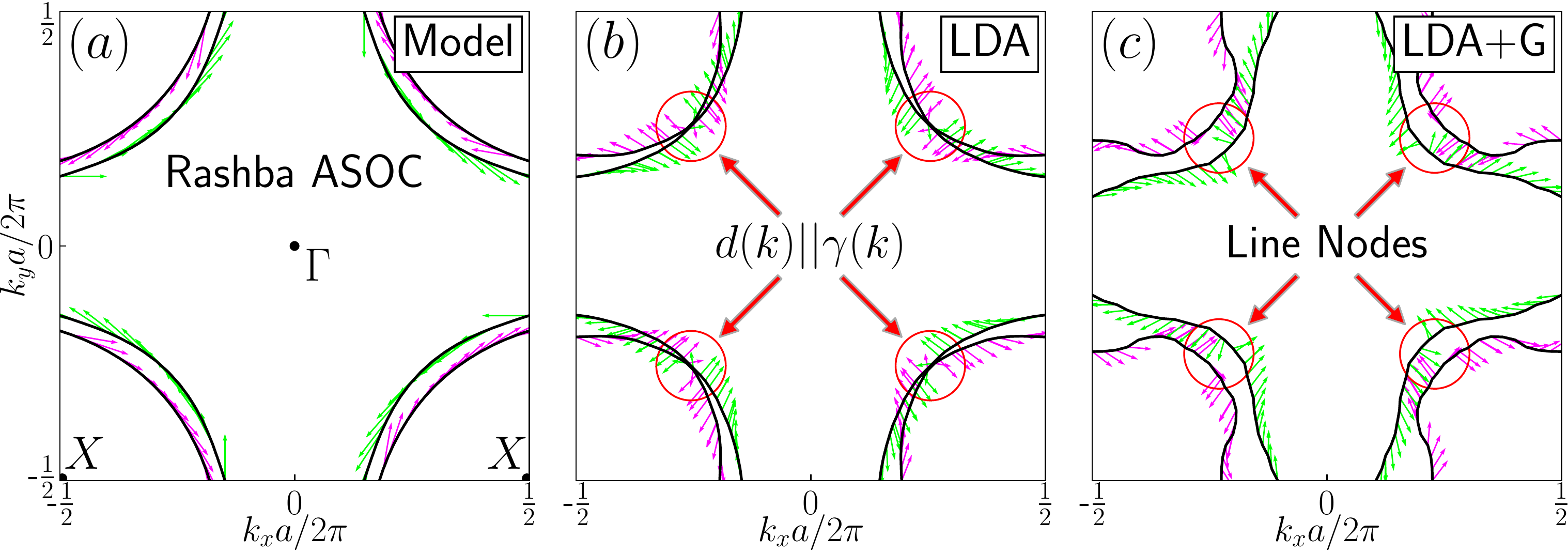}
	\caption{Plots of the FS of CeCoGe$_3$ within the
		$k_z=0$ plane for (a) the TB model, (b) LDA and (c) LDA+G. Green 
		(magenta ) arrows show the direction of spins projected into the 
		$xy-$plane for the upper (lower) band at each point. For
		LDA+G the energy is shifted by -5meV to avoid FS distortion due to
		pockets created by a set of Type-II Weyl points located just above
		$E_f$.  Red circles highlight the spin distortion caused by the NLs
		in the normal state, which indicates the existence of zeros in 
		$\boldsymbol\gamma(\vec{k})$ and implies line nodes in the SC gap function.}
	\label{spins}
\end{figure*}
\indent
Moving away from the $\Gamma-Z$ axis within the $\sigma_v$ ($\sigma_d$) mirror
plane, compatibility relations dictate that the $\Lambda_6$ and $\Lambda_7$
doublets split into bands with $\Sigma_3$/$\Sigma_4$ ($\Delta_3$/$\Delta_4$)
irreducible representation. They can be distinguished by their mirror
eigenvalue, with $-i$ corresponding to $\Sigma_3$/$\Delta_3$
and $+i$ to $\Sigma_4$/$\Delta_4$. Intersecting bands belonging to different
mirror plane irreducible representations form a topologically protected
continuous line of degeneracy called a Weyl nodal line
\cite{nl-ref}. Such NLs are protected by mirror symmetry, and are robust
against perturbations. Verification of NL topology is further discussed in the SM\cite{sup}.\\
\indent
A selection of NLs in CeCoGe$_3$ are shown in Figure \ref{nl}. In LDA, three
NLs emerge from the DP, with NL-2 and NL-3 forming loops within the
$\sigma_v$ plane and NL-4 forming a loop in the $\sigma_d$ planes. The two other NLs within the $\sigma_d$ plane, NL-1 and NL-5, do not form loops,
instead connecting across the edge of the BZ. When correlations are considered,
the NL structure of CeCoGe$_3$ changes dramatically. NL-3 mixes with other NLs
(not pictured), inverting to connect across the $k_z=0$ plane, nearly coinciding
with NL-2, while NL-4 and NL-5 are destroyed by correlations. On the other hand,
the momentum-space structures of NL-1 and NL-2 do not change much in LDA+G. We
note that since NL-3 and NL-4 are very small features and are strongly affected
by correlations, it is unlikely that they can be resolved experimentally. The
SM\cite{sup} contains the details of several additional NLs which
lie farther from $E_F$, for a total of 15 (12) NLs in LDA (LDA+G). \\
%
%
%
%
\indent
As we have mentioned previously, the renormalization of quasiparticle bands by
correlations affects not only the momentum-space position of topological
features, but also the energy at which they are located. Coulomb interactions
substantially reduce the width of the Ce-$4f$ bands and pin them to the Fermi
energy due to the increased density of states. A consequence of this
renormalization is that any topological features formed by the Ce-$4f$ bands
move closer to the $E_F$, becoming more relevant for the SC physics.\\
%
\indent
We illustrate this by showing the renormalization of the first WP in Table
\ref{weyl-table} as well as NL-1 and NL-2 (Figure \ref{renorm}).
Since the two NLs are formed from bands with a large Ce-$4f$ component,
the renormalization of these bands by correlations has a
twofold effect, narrowing the energy dispersion of the NLs and move them closer
to the $E_F$. Likewise, the Weyl point located at $(0.09700
\frac{2\pi}{a}, 0.18704 \frac{2\pi}{a}, 1.0 \frac{2\pi}{c})$ is formed from
bands that have primarily Ce-$4f$ character near this momentum. The
correlations introduced by LDA+G raise the energy by 60 meV, and shift the Weyl point to
a new momentum space position $(0.16138 \frac{2\pi}{a}, 
0.13255 \frac{2\pi}{a}, 1.0 \frac{2\pi}{c})$.\\
%
%
%
%
\indent
While SC in the CeTX$_3$ compounds has been studied extensively, the nature of the pairing state has not been settled.
There are a number of good reviews on superconductivity in non-centrosymmetric materials \cite{book, Smidman_2017, yip-review, ncs-bauer-review}, which we will briefly outline here.
%
The absence of inversion symmetry allows for an ASOC term,
\begin{equation}
H_{\text{ASOC}} = \sum_\vec{k} \sum_{\alpha\beta=\uparrow,\downarrow} 
\boldsymbol\gamma(\vec{k}) \cdot \tilde{\boldsymbol\sigma}_{\alpha\beta} 
c^\dagger_{\vec{k}\alpha} c_{\vec{k}\beta},
\end{equation}
where the Pauli matrices $\tilde{\boldsymbol\sigma}= (\tilde{\sigma}_x,\tilde{\sigma}_y,\tilde{%
\sigma}_z )$ act on the pseudospin basis states $\ket{\vec{k},\uparrow}$ and
$\ket{\vec{k},\downarrow}$, and $c^\dagger_{\vec{k}\alpha} (c_{\vec{k}\beta})$ are the corresponding creation (annihilation) operators.\\
\indent The form of $\boldsymbol\gamma(\vec{k})$ explicitly determines the local spin structure in $\vec{k}$-space. This places a constraint on the superconducting gap function $\Delta(k)$, which in general can be expanded in the basis of Pauli matrices as $\Delta(k) = [\psi(\vec{k})+\vec{d}(\vec{k})\cdot%
\tilde{\boldsymbol\sigma}]i\tilde{\sigma}_y$, with even-parity scalar
$\psi(k)$ (singlet) and odd-parity vector $\vec{d}(\vec{k})$ (triplet) components. For sufficiently strong ASOC, $\ket{\pm\vec{k},\uparrow}$ states become non-degenerate, which suppresses the component of $\vec{d}(\vec{k})$ that is not parallel to $\boldsymbol\gamma(\vec{k})$ \cite{book, Smidman_2017, Agterberg2004, samokhin-mineev}. It then follows that the triplet component of the gap $\vec{d}(\vec{k})$ can be infered directly from the spin structure at the Fermi surface.
The symmetry of the pairing gap has been studied in the context of  anti-ferromagnetic spin-fluctuations
near the SC transition \cite{spinfluct-TKF2007, spinfluct-TKF2010}. It has also been suggested that CeRhSi$_3$ and CeIrSi$_3$ may be topological
Weyl superconductors \cite{weylSC1,weylSC2}, and indeed our present study has 
identified a number Weyl nodes in the energy dispersion. However, the WPs found in our calculations are Type-II, with a hyperbolic FS that does 
not enclose the node. Their contribution to the FS topology is quite small, and
most are too far away from $E_f$ to be relevant for the SC physics, even when taking band renormalization into account.\\
\indent
Instead we focus on the effect of topological NLs found in these compounds,
which occupy a significantly larger phase space.  Figure \ref{spins} shows cross sections of the CeCoGe$_3$ FS in the $k_z=0$ plane for LDA and LDA+G, compared to a two band (TB) model (Figure \ref{spins}a) which reproduces the principal FS features of the CeTX$_3$ family \cite{spinfluct-TKF2007}, showing a realignment of the spins beyond the usual Rashba-type ASOC due to the topological nodal lines near $E_F$. In LDA, the Type-II nodal line NL-5 passes through the $k_xk_y$-plane close to $E_F$, and its strongly tilted dispersion results in hyperboloid FS sheets around the $X$ point in Fig \ref{spins}b. The spins along the surface rotate by an angle $\pi$ in the vicinity of the NL, creating a vortex-like defect which shrinks as the energy approaches the nodal line intersection. Exactly at the nodal line energy this vortex becomes vanishingly small, but the spin texture remains continuous due to the degeneracy of the bands. An animation of the spin rotations resulting from the nodal lines is included in the SM \cite{sup}.
In LDA+G (Fig \ref{spins}c), correlations shift the NL away from $E_F$, 
resulting in a gap between the Fermi surfaces, but leave the vortex-like spin defect unaffected. This spin distortion at the $\sigma_d$ planes is a direct consequence of the topological nature of the NLs, making it distinct from spin structures beyond Rashba ASOC which have been considered in other works \cite{spins1-karsten_held, spins2, spins3}.\\
\indent
It has been proposed that line nodes in the superconducting gap function could arise as a result of a topological defect in $\boldsymbol\gamma(\vec{k})$, and that such a state would be dominated by spin-triplet pairing and robust to perturbation\cite{YS-topo}. 
%
%
%
%
%
The vortex-like defects in the spin structure that arise from the topological nodal lines in the normal state of CeTX$_3$ compounds can therefore serve as a natural origin for line nodes in the superconducting gap.
This result is consistent with experiments that have found evidence of gapless line-node superconductivity in CeRhSi$_3$ and CeIrSi$_3$ \cite{ceirsi3-nmr,ceirsi3-knight,cerhsi3-SC2,cerhirsi3-exp}. Additional experiments are needed to clarify the form of the SC gap in this family of materials.\\ 
%
\indent
%
%
%
%
%
%
\indent
In summary, we have performed simulations of SC compounds in the CeTX$_3$
series with LDA and LDA+G, choosing the double counting potential in such a way
that reproduces the experimental Fermi surfaces. We characterized the
topological properties of their energy dispersion finding WPs and NLs, which
are renormalized close to the $E_F$ by the strong Coulomb interactions
of the Ce$-4f$ orbitals.  These topological features in turn affect the
spin-structure at the FS in these materials, which we have used to make a
first-principles prediction of the superconducting gap structure.\\ 
\vspace{-14pt}
\begin{acknowledgments}
This work was supported by NSF DMR Grant No. 1832728. X.G.W. acknowledges the
support from NSFC 11834006 and the Tencent Foundation through the XPLORER PRIZE.
\end{acknowledgments}
\vspace{-22pt}
\bibliography{references}
\bibliographystyle{apsrev4-1}
\end{document}